\documentclass[journal=jacsat,manuscript=article]{achemso}

\usepackage[version=3]{mhchem} 
\usepackage[T1]{fontenc}       
\usepackage{graphicx}  
\usepackage{epstopdf}  

\author{L.~Marseglia}\affiliation{Research Laboratory of Electronics, Massachusetts Institute of Technology, 50 Vassar St, Cambridge, MA 02139 USA}\email{lucamars@mit.edu}
\author{K.~Saha}\affiliation{Research Laboratory of Electronics, Massachusetts Institute of Technology, 50 Vassar St, Cambridge, MA 02139 USA}
\author{A.~Ajoy}\affiliation{Research Laboratory of Electronics, Massachusetts Institute of Technology, 50 Vassar St, Cambridge, MA 02139 USA}
\author{T.~Schr\"{o}der}\affiliation{Research Laboratory of Electronics, Massachusetts Institute of Technology, 50 Vassar St, Cambridge, MA 02139 USA}
\author{D.~Englund}\affiliation{Research Laboratory of Electronics, Massachusetts Institute of Technology, 50 Vassar St, Cambridge, MA 02139 USA}
\author{F.~Jelezko}\affiliation{University of Ulm, Helmholtzstr 16, 89081 Ulm, Germany}
\author{R.~Walsworth}\affiliation{Harvard-Smithsonian Center for Astrophysics and Dept of Physics, Harvard University, Cambridge, MA 02138 USA}
\author{J.L.~Pacheco}\affiliation{Sandia Labs,1515 Eubank SE Albuquerque, NM 87123, USA}
\author{D.L.~Perry}\affiliation{Sandia Labs,1515 Eubank SE Albuquerque, NM 87123, USA}
\author{E.S.~Bielejec}\affiliation{Sandia Labs,1515 Eubank SE Albuquerque, NM 87123, USA}
\author{P.~Cappellaro}\affiliation{Research Laboratory of Electronics, Massachusetts Institute of Technology, 50 Vassar St, Cambridge, MA 02139 USA}

\title{A bright  nanowire single photon source  based on SiV centers in diamond}

\abbreviations{NV,SiV}

\begin{document}

\begin{tocentry}
\includegraphics[width=9cm, height=3.5cm]{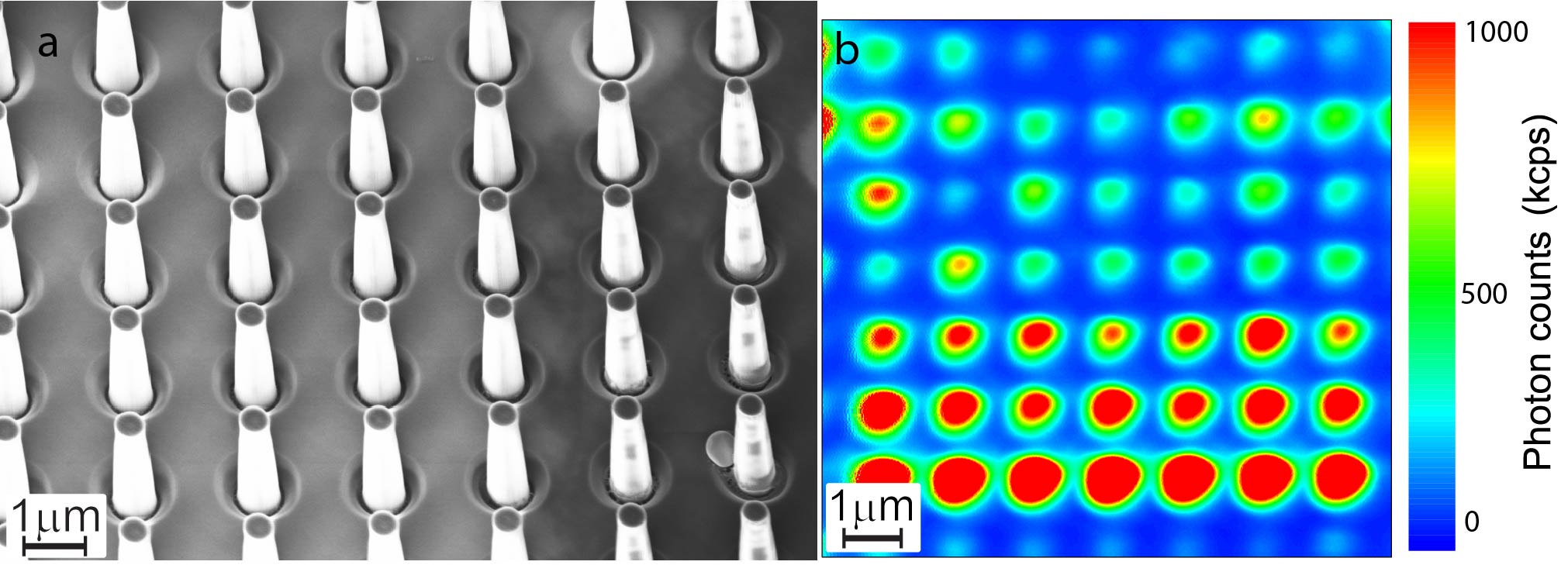}
\end{tocentry}

\begin{abstract}
The practical implementation of many quantum technologies relies on the development of robust and bright single photon sources that operate at room temperature. The negatively charged silicon-vacancy (SiV$^-$) color center in diamond is a possible candidate for such a single photon source. However, due to the high  refraction index mismatch to air, color centers in diamond typically exhibit low photon out-coupling. An additional shortcoming is due to the random localization of native defects in the diamond sample. Here we demonstrate deterministic implantation of Si ions with high conversion efficiency to single SiV$^-$ centers, targeted to fabricated nanowires. The co-localization of single SiV$^-$ centers with the nanostructures yields a ten times higher light coupling efficiency than for single SiV$^-$ centers in  bulk diamond. This enhanced photon out-coupling, together with the intrinsic scalability of the SiV$^-$ creation method, enables a new class of devices for integrated photonics and quantum science.
\end{abstract}

A practical source of  indistinguishable single photons~\cite{Obrien_09} is crucial for many quantum science applications, including optical quantum computing~\cite{Sipahigil14} and many quantum repeater protocols. An ideal single photon source must have a fast repetition rate, and also must be deterministic or ``on-demand'', with 100$\%$ probability of emitting a single photon and subsequent photons being indistinguishable. Additionally, for practical implementations a single photon source must be photostable at room temperature. Solid-state atom-like emitters are of great interest as single photon sources for practical quantum optical devices~\cite{Li_15}, because they satisfy most of the required characteristics and can be easily integrated with other nanophotonic devices in a scalable way. In the past decade, color centers in diamond have emerged as particularly promising candidates for practical single photon sources, due to their brightness and photostability at room temperature. Many defects in diamond have been investigated, including Nickel-based NE$8$ centers~\cite{Wu07}, nickel-silicon complexes~\cite{Aharonovich09}, Nitrogen-vacancy (NV) centers~\cite{Doherty13}, and chromium-related color centers~\cite{Kennard13}. The NV center is perhaps the most studied, due to its photostability and unparalleled spin properties~\cite{Doherty13} relevant to many applications~\cite{Childress14,Taylor08}. Integration of the NV center in nanophotonic structures created directly in diamond has also been demonstrated~\cite{Marseglia11,Maletinsky12,Burek14,Englund15,Piracha16}. However, NV centers possess a broad fluorescence spectrum of about $100$~nm linewidth at room temperature, with only about $4\%$ of their emission falling in the zero phonon line (ZPL), making them very challenging to be used as a source of indistinguishable single photons.

In comparison, the negatively-charged silicon-vacancy (SiV$^-$) center in diamond~\cite{Neu11} has been shown to be a more suitable candidate as a single photon source owing to its strong ZPL. The SiV$^-$ center consists of an interstitial Silicon atom between two vacancies in the  carbon lattice. It is a bright source of indistinguishable single photons with the remarkable characteristic of having  $70\%$ of its emission in the ZPL, with a width of $0.1$~nm at low temperature (4 K). The SiV$^-$ center also possesses an inversion symmetry, which leads to high spectral stability~\cite{Rogers14}, allowing creation of indistinguishable single photons from separate emitters~\cite{Sipahigil14}.
\begin{figure*}[ht]\centering
\includegraphics[width=0.95\linewidth]{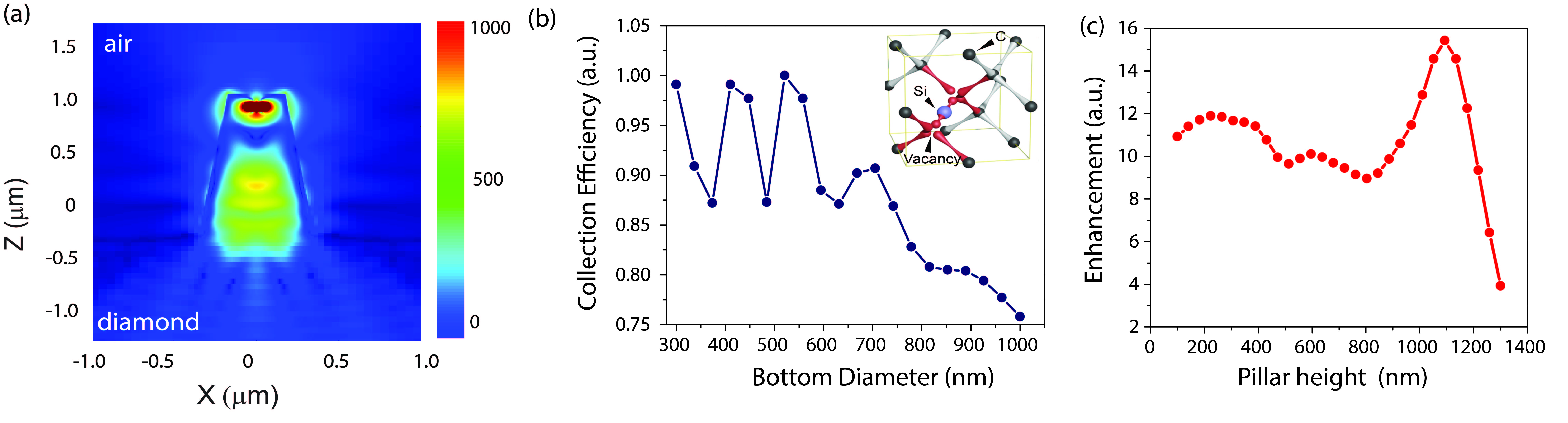}
\caption{\textbf{Single photon source based on a single SiV$^-$ center in a diamond nanowire} \textbf{(a)},  Simulation of the intensity of the optical emission from a single SiV$^-$ embedded in a nanowire. \textbf{(b)}, Simulated correlation between nanowire collection efficiency and bottom diameter, with the top diameter fixed at the optimal value ($350$~nm). Inset: SiV$^-$ atomic structure. \textbf{(c)}, Simulated correlation between enhancement due to the nanowire and position of a single photon emitter measured from the substrate, i.e., height of SiV$^-$ in nanowire.}
\label{fig:siv}
\end{figure*}
SiV$^-$ defects occur very rarely in natural diamonds, but can be purposely fabricated.
Existing strategies to create SiV$^-$ include silicon contamination during CVD growth process in diamond~\cite{Neu13} or etching structures in a hybrid diamond-silicon carbide material~\cite{Zhang16}. However, these techniques give very low control of the location of individual SiV$^-$ centers. An alternative  strategy exploits ion implantation with a Focused Ion Beam (FIB) to achieve precise spatial localization of SiV$^-$ defects~\cite{Moller14}. FIB is used at low ion intensity to create high resolution secondary electron emission images, and at high ion intensity to etch away material present on the surface of the sample. By replacing the standard gallium ion source with a silicon ion source, FIB also allows creation of SiV$^-$ centers with  position accuracy of tens of nanometers in three dimensions~\cite{Tamura14}. Here we report fabrication of monolithic nanowires ~\cite{Babinec10} coupled to single SiV$^-$ centers created by deterministic silicon ion implantation. We experimentally demonstrate that these nanostructures achieve an increase in SiV$^-$ fluorescence photon count rate by a factor of ten with respect to defects in bulk diamond; and also reduce the optical power needed to excite the color centers by an order of magnitude.

To select the optimal geometry of nanowires, given constraints in both the fabrication and defect creation, we first performed finite difference time domain (FDTD) simulations of a nanowire with a single SiV$^-$ emitter of wavelength $ \lambda = 738 $~nm embedded in it. We estimated the total emission of light coupled out in the vertical direction from the nanowire in different configurations by varying the diameter of the nanowire and the position of the emitter therein. Due to practical fabrication constraints, we chose a truncated cone structure (Fig.~\ref{fig:siv}(a)) instead of a  cylindrical structure, because less control is needed over the aspect ratio during etching. We simulated the optical collection efficiency as a function of the nanowire diameter at the bottom and top of the structure, obtaining an optimal value of $350$~nm for the top diameter and about $650$~nm for the bottom diameter (Fig.~\ref{fig:siv}(b)). We also accounted for implantation constraints that limit the maximum penetration of silicon ions in diamond given by the FIB energy to a depth of $120\pm 26$~nm. Accordingly, we simulated the emitter properties in the nanowire at a depth of $120$~nm below the surface and found that the best value for height, compatible with fabrication constraints, is around 1 $\mu$m. In Fig.~\ref{fig:siv}(c) we show that the fabricated nanowires have height close to the optimized value.

Before performing deterministic creation and characterization of SiV$^-$ in the nanowires via our implantation technique, we also tested an alternative strategy where nanowires were fabricated in a sample already containing SiV$^-$ defects. We used a sample containing a high density of SiV$^-$ color centers, which was previously characterized in a different experiment where we fabricated an array of Solid Immersion Lenses (SILs) in an area with high density of SiV$^-$~\cite{Rogers14}. While the SILs could provide an enhancement of the detected light coupling by a factor of 10, the minimum possible SIL size ($\simeq$1 $\mu m$) is not suitable for applications to integrated optical circuits. This limitation can be overcome by fabrication of nanowires that are only several hundred nanometers in diameter. For such fabrication, we used a diamond sample with high purity: a $^{12}$C-enriched CVD layer grown on a low strain, type-IIa, high-pressure high-temperature (HTHP) substrate.
During the growth process a $6$H-SiC single-crystal plate was used as a silicon source to allow doping of the diamond. Using a home-built confocal microscope, we characterized the presence of uniformly distributed SiV$^-$ color centers in a 15 $\mu$m layer. In this sample we created a large array of nanowires~\cite{Babinec10} via electron-beam lithography and reactive-ion etching (RIE) techniques (see Methods). This procedure has the distinct advantage that the implantation step after the fabrication is avoided, with the potential to create large arrays of photon emitting nanowires. However, this advantage comes at the cost of random positions of the SiV$^-$ in the nanostructure. Indeed, we found it was extremely challenging to find a single SiV$^-$ emitter in a nanowire and locate its position precisely, using this methodology. Due to the high concentration of SiV$^-$, out of 100 nanowires we found only one that contained a single SiV$^-$ emitter. Moreover, the coupling of this single SiV$^-$ to the nanowire was not as good as we expected, showing an increase in the emitted light intensity of only a factor of three compared to a single SiV$^-$ in a bulk.
We attribute this poor performance to a deeper location of the SiV$^-$ in the diamond, below the nanostructure, or to a non-optimal position of the SiV$^-$ in the structure. While a possible solution would be to repeat this fabrication strategy on a sample with very low concentration of SiV$^-$, this approach would void its main advantage (the simultaneous creation of many nanowires containing photon emitters) since it would lead to a very low yield in the realization of a nanostructure coupled to single SiV$^-$.

\begin{figure*}[ht]\centering
\includegraphics[width=0.9\linewidth]{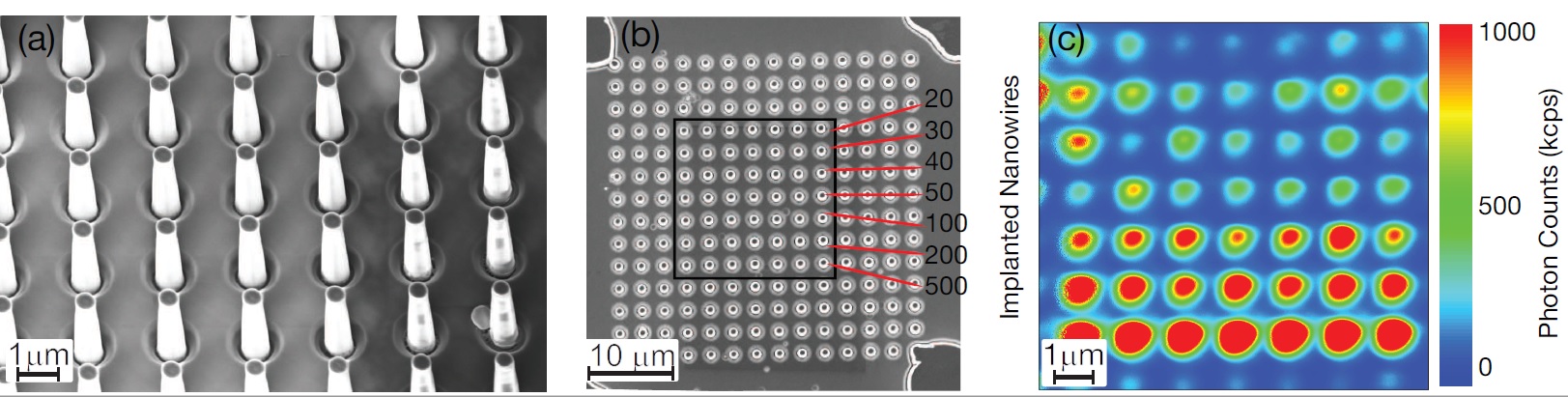}
\caption{\textbf{Array of diamond nanowires containing SiV$^-$ color centers} \textbf{(a,b)} Secondary electron emission images of the array of diamond nanowires studied in the present work.  Each nanowire has a $350$~nm top diameter, $650$~nm bottom diameter, and height of 1.2 $\mu m$ . \textbf{(c)} Confocal microscopy scan (15 $\times$ 15 $\mu m^2$) of nanowire array after SiV$^-$ creation with implantation dose of 20 ions/4.9ms. The first three rows of the array contain respectively 10(first row), 7(second row)  and 5 (third row) SiV$^-$ centers per nanowire, while the other rows contain single or double SiV$^-$ per nanowire. Measured SiV$^-$ fluorescence count rate shows a clear correlation between the number of Si ions implanted and the number of SiV$^-$ created. Pixel size is $200$~nm; integration time per pixel is 50 ms; point spread function is $PSF\simeq 1 \mu$m; and excitation laser power is $P_{Laser}=10$ mW.}
\label{fig:array}
\end{figure*}

Given these results, we therefore pursued an evolved version of the procedure introduced earlier: by first fabricating the photonic structures and then creating defects in a quasi-deterministic manner, we were able to achieve a high yield of nanowires containing single SiV$^-$ centers. We used electron-beam lithography and RIE techniques to create a large array of nanowires (Fig.~\ref{fig:array}(a)) in a pure type-IIa, HPHT diamond crystal without SiV$^-$. Then, the image file of the fabricated arrays was used as a  map (Fig.~\ref{fig:array}(b)) to deterministically implant silicon ions~\cite{Evans15}. Focused ion implantation was performed at the Ion Beam Laboratory (IBL) at Sandia National Labs using the nanoImplanter (nI). The nI is a $100$ kV focused ion beam (FIB) machine (A\&D FIB$100$nI) with a three-lens system that is designed for high mass resolution using an E$\times$B filter and single ion implantation using fast beam blanking with beam diameters on target between $10-50$ nm depending on ion energy. The E$\times$B filter has a resolution of $> 61$ (M/$\Delta$M) allowing the user to select the ion, isotope, and energy via a variety of liquid metal alloy ions sources (LMAIS), allowing for ion beams from $\sim 1/3$ of the periodic table over a range of energies from $10$ to $200$ keV.  The implanted ion dose can be controlled down to the single ion level using a fast blanking and chopping system (minimum pulse width of 16 ns) or the external beam blanker.  The nI is a direct write lithography platform combining a Raith patterning system with a laser interferometry-driven stage.  The ion positioning is limited by the beam spot size on target convolved with the longitudinal and lateral ion straggle in the substrate.  For implantations reported here, we used a AuSbSi source to generate the Si beam. Typical beam currents range from $0.4$ to $1$ pA with a spot size of $<40$ nm. Silicon ions were implanted with an energy of $200$~keV, giving an estimated depth penetration in the nanowire of $120$~nm. The dose of silicon ions was varied for each row of the nanowire array, from $20$ to $500$ ions per spot, with lateral resolution of $30$~nm. High vacuum annealing was performed after the implantation process to repair lattice damage and to facilitate the motion of  vacancies toward the implanted silicon ions,  leading to the creation of single and multiple SiV$^-$ centers embedded in the nanowires. The implantation doses were  chosen to create from $1$ to  $8$ SiV$^-$ per nanowire upon annealing above 800 $C^\circ$, with an estimated conversion efficiency of $5\%$. We implanted a range of doses in order to obtain SiV$^-$ densities ranging from small ensembles to single color centers in different rows of the pillars. This tailored fabrication procedure is advantageous in providing a deterministic way to create a single SiV$^-$ in the desired location, in this case within a nanowire. The main disadvantages of this approach, besides involving an additional implantation step, is that the depth of the resulting SiV$^-$ defect is directly determined by the implantation energy. Still, from our simulations we found that having an emitter close to the nanowire top surface, instead of at its center, decreases the light collection emission only by a factor of $15\%$.

\begin{figure*}[ht]\centering
\includegraphics[width=0.9\linewidth]{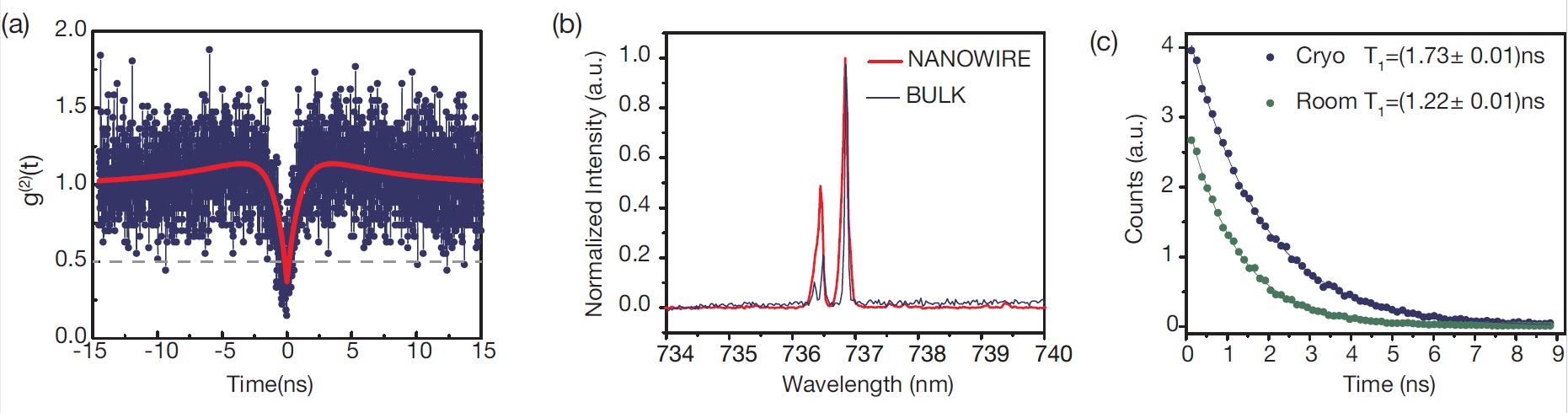}
\caption{\textbf{Characterization of single SiV$^-$ centers in nanowire} \textbf{(a)}, Second order autocorrelation function $g^{(2)}(\tau)$ fitting (in red) the raw coincidence rate of measured single SiV$^-$ fluorescence from one nanowire, with time bin $\delta = 0.012$ ns, total integration time of $10,000$ s, and laser excitation power $P_{Laser}=1$ mW. \textbf{(b)} Spectra comparison for fluorescence from a single SiV$^-$ in bulk diamond with linewidth $\Delta \lambda = 0.24$ nm, and a single SiV$^-$ in a nanowire with linewidth $\Delta \lambda = 0.1$ nm, demonstrating SiV$^-$ spectral stability after implantation. A small third spectral feature from the bulk SiV$^-$ results from reduced overlap of the C-line and D-line due to strain in the bulk diamond. \textbf{(c)} Lifetime measurements for a single SiV$^-$ in a nanowire at room and low (4 K) temperature, giving $T_1$ values of, respectively, $T^{RT}_{1}=1.22$ ns and $T^{LT}_{1}=1.73$ ns.}
\label{fig:ab}
\end{figure*}

To characterize the resulting SiV$^-$ defects, we measured non-classical light emission using a home-built confocal microscope with a Hanbury-Brown Twiss (HBT) interferometer to record the second order autocorrelation function $g^{(2)}(\tau)$. In Fig.~\ref{fig:array}(c) we show a confocal microscopy raster scan of the implanted array, showing varying intensity of the light collected from the nanowires. In the bottom part of the scan we have the highest dose of implanted silicon ions ($500$ Si per spot), which leads to a large ensemble of SiV$^-$ embedded in the nanowires and a very high  intensity of the light emission.
In the upper part of the scan, the lowest implantation dose ($20$ Si per spot) gave rise to single SiV$^-$ creation in the nanowires. $g^{(2)}(\tau)$ measurements of these single SiV$^-$ defects showed the typical antibunching behavior of a single photon emitter (Fig.~\ref{fig:ab}(a)).
To evaluate these single SiV$^-$ $g^{(2)}(\tau)$ results, the raw correlation data $c(\tau)$ was normalized and corrected in the following way. The raw coincidence rate $c(\tau)$, collected by sweeping the time $\tau$ from $0$ to $T$ in steps of $\delta$, was first normalized  according to~\cite{Brouri00,Beveratos02,Berthel15}:
\begin{equation}
g^{(2)}_{exp}(\tau)=\frac{c(\tau)}{N_1N_2 T\delta},
\label{Cn}
\end{equation}
where $N_{1,2}$ are the count rates at each detector. The normalized coincidence function $g^{(2)}_{exp}(\tau)$  differs from that of the theoretical second order autocorrelation function $g^{(2)}(\tau)$  of a single emitter according to the formula
\begin{equation}
g^{(2)}_{exp}(\tau)= g^{(2)}(\tau) \rho^2 + (1-\rho^2),
\label{g2_raw}
\end{equation}
where $\rho=\frac{S}{(S+B)}$ was calculated from the signal ($S$) and background ($B$) rates, which were determined independently during the measurement of the photon flux from a single SiV$^-$ in a nanowire. In particular, the background is the count rate measured from an empty nanowire. With this correction we could correctly determine the depth of the antibunching dip, thus properly identifying single emitters.
Finally, following the three-level system model discussed in ~\cite{Rarity98,Neu11,Neu13} we fit the data with the function:
\begin{equation}
g^{(2)}_{exp}(\tau) = (1-(1+a)e^{-\frac{|\tau|}{\tau_1}}+a e^{-\frac{|\tau|}{\tau_2}})\rho^2 + (1-\rho^2) .
\label{g2}
\end{equation}
This fit includes three parameters $a$, $\tau_1$, and $\tau_2$ that determine the characteristics of the $g^{(2)}$ function: $\tau_1$ governs the antibunching dip, whereas $\tau_2$ gives the time constant of the bunching, and the parameter $a$ determines the magnitude of the bunching.
We demonstrated the creation of single photon SiV$^-$ emitters in correspondence to the lowest implantation doses ($20$ and $30$ ions per spot), by observing  the autocorrelation function minimum $g^{(2)}(\tau)<0.5$. In the nanowires with implantation doses corresponding to a value of $30$ and $40$ ions per spot, we were also able to measure values of the autocorrelation function $0.5 < g^{(2)}(\tau)<0.7$, indicating the presence of a double emitter.

Moreover in Fig. \ref{fig:ab}(c) we show  the measured spectra at temperature $T=4$ K for a single SiV$^-$ in the nanowire and a single SiV$^-$ created by the same technique but in a bulk sample. This measurement shows an almost perfect spectral overlap between the two emitters, which leads to the conclusion that SiV$^-$ implantation in the nanowires does not affect the spectral stability of the SiV$^-$. We also measured the lifetime of the excited state of a single SiV$^-$ in a nanowire at room temperature and at low temperature (4 K) as shown in Fig. \ref{fig:ab}(d). The radiative decay lifetime of SiV$^-$ increases from 1.22 ns at room temperature to 1.7 ns at 4 K, corresponding to a transform-limited PLE linewidth of 94 MHz. These values are in good agreement with the values for bulk SiV$^-$ shown in previous work~\cite{Rogers14}.

\begin{figure*}[hbpt]\centering
\includegraphics[width=0.9\linewidth]{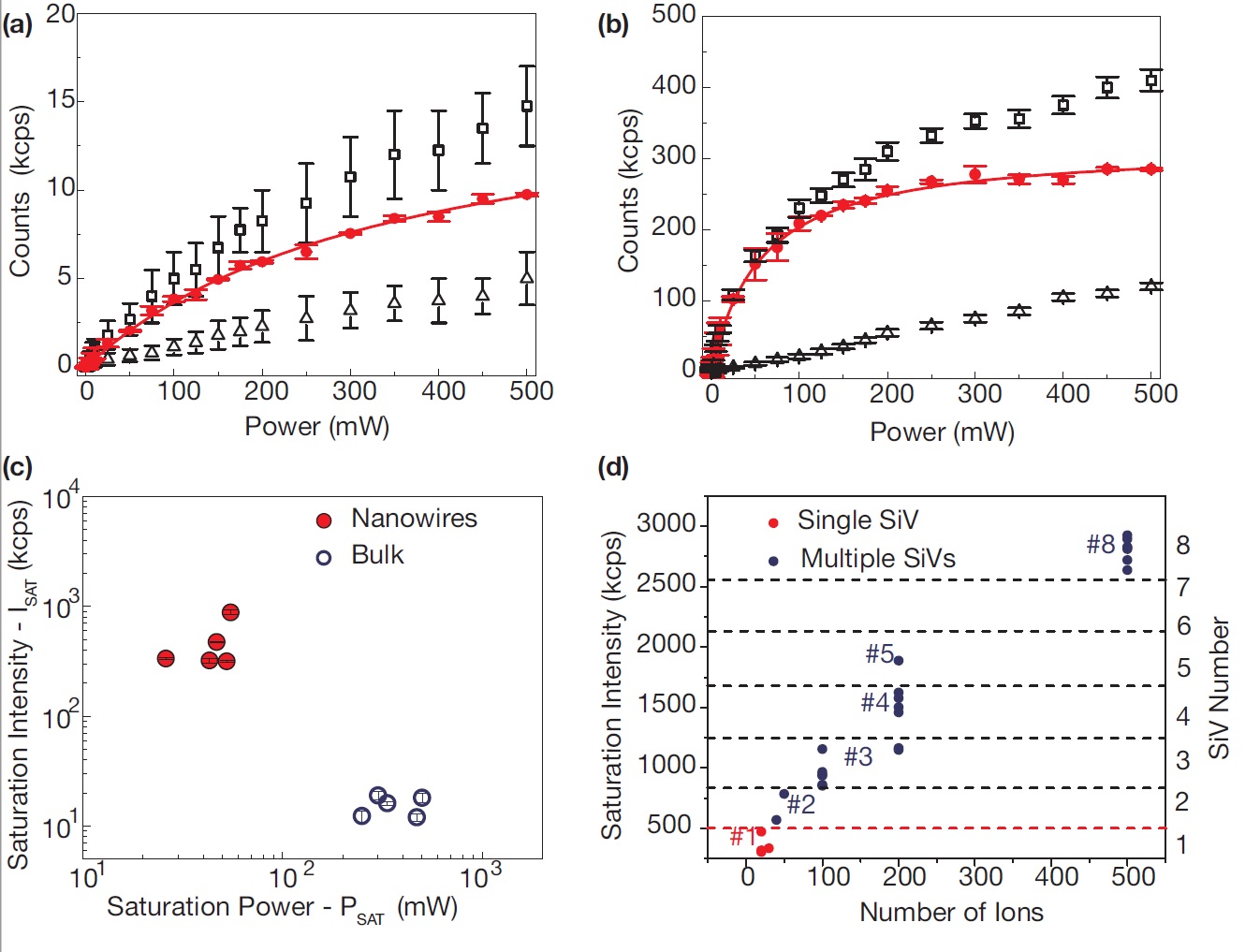}
\caption{\textbf{Single SiV$^-$ fluorescence rates in bulk diamond and nanowires.} Measured fluorescence rates as a function of laser excitation power for single SiV$^-$ centers in \textbf{(a)} bulk diamond and \textbf{(b)} a nanowire. Open black squares are raw counts; hollow black triangles are background; and filled red circles are normalized SiV$^-$ counts. Backgrounds were measured, respectively, in an empty zone of the bulk with no SiV$^-$, and in a nanowire not implanted with silicon ions. \textbf{(c)} Single SiV$^-$ optical properties ($P_{sat}$ and $I_{sat}$)) for several nanowires in the array (red circles) compared to example single SiV$^-$ centers in the bulk (blue open circles).\textbf{(d)} Estimated number of SiV$^-$ defects created per nanowire, determined from measured values of $I_{sat}$. Red circles indicate single SiV$^-$ centers. }
\label{fig:saturation}
\end{figure*}

A crucial benefit of  creating these nanostructures coupled to  SiV$^-$ emitters is the increased light out-coupling, as can be observed by comparing the photon flux from an individual SiV$^-$ embedded in the nanowire to an individual SiV$^-$ emitter in the bulk crystal. We measured the count rate of photons emitted by  single SiV$^-$ centers as a function of the power of the excitation laser. Fig.~\ref{fig:saturation}(a,b) shows the saturation curves, respectively, for a single SiV$^-$ in bulk and a single SiV$^-$ in the nanowire. After a rise at low pump powers $P$, the intensity $I$ of the light emitted from a single SiV$^-$ saturates at a value ($I=I_{sat}$) corresponding to the saturation power  ($P=P_{sat}$). This bound in the emission rate is  due to the quantum nature of the photon emitter, set by  the SiV$^-$ spontaneous decay rate. The emission intensity is a function of the laser power according to the formula~\cite{Neu11}
\begin{equation}
I(P)= I_{sat} \frac{P}{P+P_{sat}}.
\label{sat}
\end{equation}
We  fitted the saturation curves for seven single SiV$^-$ centers coupled to nanowires and  seven single SiV$^-$ centers in bulk diamond to extract the  saturation power and saturation intensity. The resulting values (Fig.~\ref{fig:saturation}(c)) show that the average saturation intensity for a single SiV$^-$ in a nanowire is at least one order of magnitude higher than for a single SiV$^-$ in the bulk; and the power needed to excite a single SiV$^-$ emitter in a nanowire is also at least an order of magnitude lower than for a single SiV$^-$ in the bulk. Specifically, we obtained an average saturation intensity for a single SiV$^-$ in a nanowire of $\langle I_{sat} \rangle =355$~kcps. We further used this value to estimate the number of SiV$^-$ in individual nanowires containing more than one or two emitters. The results are shown in Fig.~\ref{fig:saturation}(d), where one can observe a clear correlation between the number of silicon ions implanted in the nanowires and the estimated SiV$^-$ defect number, thus demonstrating that it is possible to create a desired number of SiV$^-$ defects with very high accuracy.

In conclusion, we  report the fabrication of bright sources of single photons from deterministically implanted SiV$^-$ defects coupled to fabricated nanowires in diamond. We also demonstrated an improvement in the photonic coupling of  single SiV$^-$ emitters, thanks to the presence of the nanowire. SiV$^-$ in nanowires are pumped ten times more efficiently than single SiV$^-$ in the bulk and allow ten times higher single photon count rates. This result will enable a new class of devices for integrated photonics and quantum information processing. One of the major advantages offered by SiV$^-$ compared to other solid-state emitters arises from the inversion symmetry of the SiV$^-$ defect, which leads to a stable single photon source immune to spectral diffusion, making the SiV$^-$ an ideal candidate for integration into diamond nanophotonic structures. Here, we showed that this key characteristic is preserved for SiV$^-$ defects implanted in photonic nanostructures.
These results pave the way for more complex experiments, e.g., exploiting quantum interference from multiple SiV$^-$ embedded in nanostructures.

\begin{acknowledgement}
The authors thank Tokuyuki Teraji and Junichi Isoya for providing the diamond sample, Yuliya Dovzhenko and Tony X. Zhou for discussions of diamond etching techniques, and Alp Sipahigil and Mikhail Lukin for discussions and valuable insight regarding SiV$^-$ physics and applications. Financial support was provided in part by  NSF grant No. PHY1125846 (at MIT), the Army Research Laboratory Center for Distributed Quantum Information (CDQI), in part by the NSF EFRI-ACQUIRE program "Scalable Quantum Communications with Error-Corrected Semiconductor Qubits", and in part by the U. S. Army Research Laboratory and the U. S. Army Research Office under contract/grant number W911NF1510548 (at Harvard). This work was performed in part at the Center for Nanoscale System (CNS) of Harvard University. This work was also performed in part at the Center for Integrated Nanotechnologies, an Office of Science User Facility operated for the U.S. Department of Energy (DOE) Office of Science. Sandia National Laboratories is a multi-program laboratory managed and operated by Sandia Corporation, a wholly owned subsidiary of Lockheed Martin Corporation, for the U.S. Department of Energy's National Nuclear Security Administration under contract DE-AC04-94AL85000.

\end{acknowledgement}

\begin{suppinfo}

($1$)Device Fabrication and ($2$) Measurement Setup.

\end{suppinfo}

\bibliographystyle{unsrt}
\bibliography{Marseglia_17}

\end{document}